\documentstyle[aps,prl,multicol,epsf]{revtex}

\def\e{\epsilon}

\def\s{\sigma}

\def\D{\Delta}

\def\O{\Omega}

\def\IR{\relax{\rm I\kern-.18em R}}
\font\cmss=cmss10 \font\cmsss=cmss10 at 7pt
\def\IZ{\relax\ifmmode\mathchoice
{\hbox{\cmss Z\kern-.4em Z}}{\hbox{\cmss Z\kern-.4em Z}}
{\lower.9pt\hbox{\cmsss Z\kern-.4em Z}}
{\lower1.2pt\hbox{\cmsss Z\kern-.4em Z}}\else{\cmss Z\kern-.4em Z}\fi}
\def\IN{\relax{\rm I\kern-.18em N}}

\input epsf
\begin{document}
%\draft command makes pacs numbers print
\draft
\widetext
\title{Interplay of spin density wave and superconductivity with
different pairing symmetry}
\author{Haranath Ghosh}
\address{Department of Physics, University of Arizona, Tucson, AZ 85721, USA.}
\author{S. Sil}
\address{Theoretische Tieftemperaturephysik, Gerhard Mercator
Universit\"{a}t, Duisburg 470 48, Germany.}
\author{S. N. Behera}
\address{Institute of Physics, Sachivalaya Marg, Bhubaneswar -751 005, INDIA.}

\maketitle
\begin{abstract}
A model study for the coexistence of the spin density wave and
superconductivity is presented.
With reference to the recent angle
resolved photo emmission experimental data in high $T_c$ cuprates,
presence of the nested pieces of bands is assumed.
The single band Hubbard model, therefore,
when treated within the Hatree-Fock mean field theory leads to
a spin density wave (SDW) ground state. The superconductivity (SC)
is assumed
to be due to a generalised attractive potential with a separable form
without specifying to any particular origin. It therefore allows a
comparative study of the coexistence of superconductivity of different
order parameter symmetry with the spin density wave state. We find that
the phase diagram, comprising of the amplitudes of the respective gaps
(SC and SDW) Vs. band filling resembles to that of the high $T_c$ cuprates
only when the order parameter of the superconducting phase has $d$-wave
symmetry. Thermal variation of different order parameters (e.g, SC and SDW)
also show interesting coexistence and reentrance behaviors that are consistent
with experimental observations, specially for the borocarbides.

\end{abstract}
\begin{multicols}{2}

\narrowtext
\tightenlines
\section{Introduction}
\label{sec:intro}
The coexistence of magnetism and superconductivity has intrigued both
experimentalists and theoreticians for a long time.
It is well known that the superconductivity (SC) and
spin density wave (SDW) co-exist in a large classes of systems e.g.,
organic layered superconductors \cite{1}, the Bechgaard salts $(TMTSF)X$,
with $X=PF_6$, $AsF_6$, $ClO_4$ etc. \cite{2}, heavy fermion systems \cite{3},
high temperature superconductors \cite{4} and recently discovered borocarbides
\cite{5}.

\par In case of the organic superconductors belonging to the family
$(TMTSF)_2X$ ($TMTSF \equiv $ tetramethyltetraselena fulvalence and $X=PF_6$,
$AsF_6$ etc.) show coexistence of superconductivity and spin density wave
like antiferromagnetism (SDW) at low temperature in moderate pressures
($\sim$ 7 Kbar) \cite{6}. Many compounds ($X=PF_6$, $AsF_6$, $SbF_6$ etc.)
exhibit metal - insulator transition at 12 - 17 K under ambient pressure
which is caused by spin density wave (SDW) transition. This transition
is suppressed by moderate pressures that results in superconductivity.
However, the nature of this coexistence is very much different
from what would encounter in case of ternary compounds. In organic
superconductors, there is only one conduction band and the
antiferromagnetic ordering destroys the Fermi surface (FS) due to its $2k_F$
periodicity which results in an insulating phase. In high-$T_c$ cuprates also
due to the low dimensionality aspects there exists similar periodicity
due to nesting of the FS \cite{7}, and leads to FS instability by forming
the SDW state.
On the other hand, in intermetallic compounds the magnetism is due to
localized $f$ electrons and superconductivity arises due to the
conduction electrons. The organic superconductors in turn, are essentially
quasi one dimensional compounds with weak interaction between the adjacent
chains. Our main concern in the present paper will however be to discuss the
coexistence of antiferromagnetism (SDW) and superconductivity (SC) in the
cuprates.
\par
In the high temperature cuprate superconductors $\rm La_{2-x}(Sr,Ba)_xCuO_{4-y}$
and
$\rm YBa_2Cu_3O_{7-y}$ strong antiferromagnetic spin order with 
a commensurate wave
vector $\rm Q=(\pi/a,\pi/b)$ has been observed in neutron \cite{8} and
Raman scattering \cite{9}
experiments.
Moreover, in contrast to the conventional low $T_c$ superconductors, the
coherence length of high-$T_c$ materials is short (10 to 20 $\AA$) ;
the average magnetic field seen by the cooper pairs need not be zero.
Thus it appears that
the antiferromagnetism and superconductivity are intimately related in
these materials,
in contrast to the magnetism opposing superconductivity in conventional
superconductors.
\par
In the context of the influence of spin correlation on superconductivity in
oxide systems there exists two limiting points of view ;
$(i)$ localized picture where
$U \gg W$ and $(ii)$ itinerant picture $U \leq W$ ; where $U$ is the
strength of the onsite coulomb repulsion and $W$ is the band width.
In the large $U$ limit, the valence electrons are localized on
each $Cu$-site. To lowest
order in $t/U$ there is an antiferromagnetic exchange interaction $J$ of
the order of $t^2/U$, which produces spin ordering ($t$ being the
hopping integral) \cite{10}. In the second point of view, $U$
is assumed to be small so that the Mott
Hubbard transition  does not takes place and Bloch states form
a convenient basis
to describe the system. Now for a perfect square lattice and at half-filling,
the bands get nested with certain nesting wave vector $\vec Q$ that determines
a new super periodicity leading to spin density
wave ground state which may or may not be comensurate with the
underlying lattice.
 The spin
density wave ground state causes an energy gap $2 \Delta_{SDW}$ at the Fermi 
surface of the
undoped material, leading to an insulator. Here we shall confine our selves to
the latter point of view (i.e. $U \ll W$).
\par The SDW ground state is obtained from the single band Hubbard model
within the Hatree-Fock approximation and assuming that the nesting of the
FS  exists only in certain directions of the FS. The directions of the FS
where  nesting exists will be instable with respect to the SDW formation
whereas the superconducting instability may occur in the rest of the part
of the FS, provided there exists some attractive interaction between the
quasi-particles mediated by some boson exchange. Now, as regards the high-$T_c$
cuprates are concerned the knowledge of the exact mechanism for the high
temperature superconductivity is yet to be known. In confirmity with the
role of electronic correlations, the parent (undoped) cuprate materials
are always antiferromagnetic (SDW state in our case) but the long range
AF order vanishes sharply on doping the system. However, there is some
evidence that the AF fluctuations exists even in the superconducting phase.
Hence there have been proposals among the others that the pairing might
as well arise due to the exchange of quanta of AF fluctuations \cite{11}.
In that case, the superconducting order parameter will have a $d$-wave
symmetry. On the other hand, the SC in the system may even be caused by the
quanta of the fluctuations of the collective (amplitude and phase) modes
of the SDW state, present in our model, as proposed by Schrieffer et al.,
\cite{12} and others \cite{13} orelse the SC may even be due to normal
phonon mediated mechanism. In either of these cases the nature of the
superconducting order parameter symmetry could be different depending
on the nature of the origin of the pairing mechanism. Furthermore, there are
experimental indications that the symmetry of the order parameter of the
high-$T_c$ superconductors could be an admixture of $s$ and $d$- wave like
(i.e, $s\pm d$ or $\rm d + is$) \cite{14}. 
Therefore, we present a model study for
the coexistence of SDW with SC where the SC is caused by some generalised
attractive potential with separable form (without specifying any
particular mechanism) leading to different order parameter symmetry. It
therefore incorporates, two competing physical processes involving the
electron-hole (SDW) like pairing of opposite spins with a net momentum 
difference
($\vec Q$) between the conjugates and electron-electron (SC) pairing
of opposite spins with zero total momentum. Such a model study can also
provide an important understanding regarding the effect of the presence of
antiferromagnetic (AF) order on the pairing symmetry of the
superconducting state, which is extremely
important for high-$T_c$ cuprates as well as the recently discovered
borocarbides. Our main concern therefore boils down to study of the phase
diagram, comprising of the gap magnitudes of the respective gaps e.g,
the SDW and the SC gaps as a function of band filling (hole concentration).
Interestingly enough, we obtain the phase diagram which resembles
to that of the high-$T_c$ superconductors only when the superconducting
order parameter has the $d$-wave symmetry.
\par The rest of the paper is arranged as follows. In section II we provide
a brief mean field theory of the SDW ground state from a single band Hubbard
model. Section III contains the mean field model Hamiltonian for the
coexistence of the SDW and SC together with the derivation of the respective
gap equations. Section IV is devoted to discuss the results obtained from our
model calculatons and finally we conclude in section V.
\section{Spin density wave state}
The SDW state is a kind of antiferromagnetic state with the electronic spin
density forming a static wave. The density varies perpendicularly
as a function of position with no net magnetisation in the entire volume.
The SDW transition occurs when the spatial spin density modulation is
due to delocalization or initinerant electrons rather than the localized
one. Usually in the normal state the density $\rho_{\uparrow} (\vec r)$
of electron spins polarized
upward with respect to any quantization axis is completely cancelled by
$\rho_{\downarrow}$ of downward polarized spins.
In the SDW state, however, the
difference $\sigma(\vec r)=\rho_{\uparrow}(\vec r)
- \rho_{\downarrow}(\vec r)$ is finite and
modulate in space as a function of the position vector 
$\vec r$ in the SDW state. Such tendency of forming SDW
ground state takes place when a system possesses nested pieces of FS together
with intermidiate coulomb correlation. In the case of the SDW transition it is
the wave vector dependent static magnetic susceptibility which develops a
singularity at $\vec q~=~ \vec Q$ {\it i.e.}, $\chi(q,\omega) |_{\vec q~=~ \vec
Q \\,
\omega=0}~=~<< S^{+}(q,t)~;~S^{-}(-q,t)>>_{\omega} |_{q=Q,\omega=0} \rightarrow
\infty $ where $S^{\pm}$ are the spin raising and lowering operators and
$\vec Q$ is known as the nesting wave vector which determines the
periodicity of the SDW. This singularity in the magnetic susecptibility is
an artifact of the nesting property of the FS given by
\begin{equation}
\epsilon_{\vec k}~=~-~\epsilon_{\vec k + \vec Q}
\end{equation}
In addressing the possibilty of the coexistence of SDW and SC ordering in
$CuO_2$ based high-$T_c$ material we will look for a model of a single $CuO_2$
layer and neglect the inter-layer hopping because inter-layer hopping is weak
compared to the intra-layer hopping integral. The simpliest model one can use
to describe the antiferromagnetic ordering of the two dimensional correlated
system is the Hubbard model on a square lattice.
%Now, the minimal model for the correlated systems like the cuprates
%where the electroni correlations are taken care properly
%is best described by the Hubbard model  and is given by,
\begin{equation}
H~=~-\sum_{<ij>\sigma}~t_{ij}~C^{\dagger} _{i\sigma}C_{j\sigma}~+~U~\sum_{i}
\hat n_{i\uparrow} \hat n_{i\downarrow}
\end{equation}
where $t_{ij}$ is hopping integral between the different orbitals $i\&j$,
$ \hat n_{i\sigma}~=~C^{\dagger} _{i\sigma} C_{j\sigma} \equiv $ 
number operator
and
U is the intratomic coulomb repulsion or the cost in energy to put two
electrons on the same site. 
The interaction term in
the Hubbard Hamiltonian may be treated in the
mean field approximation to get the SDW state as follows,
\begin{mathletters}
\label{allequations} % notice location
\begin{eqnarray}
H_{I} & = & U\sum_{i}n_{i\uparrow}n_{i\downarrow}
=U\sum_{i}C^{\dagger}_{i\uparrow}
C_{i\uparrow}C^{\dagger}_{i\downarrow}C_{i\downarrow} \nonumber \\
&&
\approx U\sum_{i} [<C^{\dagger}_{i\uparrow}C_{i\uparrow}>C^{\dagger}_{i
\downarrow}C_{i\downarrow}+
C^{\dagger}_{i\uparrow}C_{i\uparrow}
<C^{\dagger}_{i\downarrow}C_{i\downarrow}> \nonumber \\
& &
-<C^{\dagger}_{i\downarrow}C_{i\uparrow}>C^{\dagger}_{i
\uparrow}C_{i\downarrow}-<C^{\dagger}_{i\uparrow}C_{i\downarrow}>
C^{\dagger}_{i\downarrow}C_{i\uparrow}] \label{equationa}
\end{eqnarray}
\begin{eqnarray}
&=&\sum_{i} [Un_{i}(C^{\dagger}_{i\uparrow}C_{i\uparrow}+
C^{\dagger}_{i\downarrow}C_{i\downarrow}) \
+\Delta^{z}_{i}
(C^{\dagger}_{i\uparrow}C_{i\uparrow}-C^{\dagger}_{i\downarrow}C_{i\downarrow})
\nonumber \\
&&
+\Delta^{+}_{i}C^{\dagger}_{i\downarrow}C_{i\uparrow}+\Delta^{-}_{i}
C^{\dagger}_{i\uparrow}C_{i\downarrow}]\label{equationb}
\end{eqnarray}
\end{mathletters}
where
%\begin{eqnarray}
$n_{i} ~=~ \frac{1}{2}~(<C^{\dagger}_{i\uparrow}C_{i\uparrow}>
+<C^{\dagger}_{i\downarrow}C_{i\downarrow}>),~~~$
$\Delta^{z}_{i} =-~\frac{U}{2}
(<C^{\dagger}_{i\uparrow}C_{i\uparrow}>~-~
<C^{\dagger}_{i\downarrow}C_{i\downarrow}>) = -U<S_z>$,
$~~\Delta^{+}_{i} = -~U~<C^{\dagger}_{i\uparrow}C_{i\downarrow}>~=-~U~<S^{-}_{i}
>$
and $\Delta^{-}_{i} ~=~ -~U~<C^{\dagger}_{i\downarrow}C_{i\uparrow}>~=~-~U~<S^{-
}_{i}>$
%\end{eqnarray}
\\

Therefore, the meanfield Hubbard Hamiltonian can be written as,
\begin{equation}
H=-\sum_{ij\sigma}t_{ij}C^{\dagger}_{i\sigma}
C_{j\sigma}+U\sum_{i}n_{i}\hat n_i+\sum_{i}\bar \sigma_{i}.\bar B_i
\end{equation}
where $ \hat n_i=C^{\dagger}_{i\uparrow}
C_{i\uparrow}+C^{\dagger}_{i\downarrow}C_{i\downarrow} $ and
$ B^{z}_{i}~=~\Delta^{z}_{i}~=~-~U~<S^{z}_{i}>~\equiv~$ the order parameter for
longitudinal magnetization, $ B^{\pm}_{i}~=~\Delta^{\pm}_{i}~=~-~U~<S^{\pm}_{i}
>~\equiv~$ transverse magnetization and
$ \bar \sigma_{i}~=~\left (~C^{\dagger}_{i\uparrow}~C^{\dagger}_{i\downarrow}
~\right )~\bar \tau~
\left ( \begin{array}{c} C_{i\uparrow}  \\ C_{i\downarrow} \end{array} \right )
$, where $\bar \tau~\equiv~$ Pauli matrices. The $2^{nd}$ term of equation (4)
corresponds to total charge of the system whereas the $3^{rd}$ term is the
total spin. Hence, eqn. (4) demonstrates that the charge and spin degree
of freedom of the Hubbard model is seperated. Now, the SDW state can be
described equivalently in terms of either the longitudinal or transverse
spin polarization. For example the SDW state with wave vector $\vec Q$,
having longitudinal spin polarization is
\begin{eqnarray}
<S^{z}_{i}>=S^{z}_{o}\times\left \{ \begin{array}{c} \cos(\vec Q.\vec R_i)
~\rm for~
~SD~on~sites.
\\
\sin(\vec Q.\vec R_i)\rm~for~SD~on~bonds. \end{array} \right .
\end{eqnarray}

In momentum representation the mean field Hamiltonian for a transverse
SDW state becomes
\begin{eqnarray}
H_{SDW}& =& \sum_{k,\sigma} (\e_k -\mu) C_{k,\sigma}^\dagger C_{k,\sigma} + \Delta
_{SDW} \sum_k
(C_{k+Q\uparrow}^\dagger C_{k\downarrow} \nonumber \\
&&
+ h.c)
\end{eqnarray}
where $\Delta_{SDW} = -U \displaystyle \sum_{k} 
<C_{k\uparrow}^\dagger C_{k-Q\downarrow}>$,
the order parameter for the transverse SDW state. The quasi particle
energy spectrum of the SDW state can be writen as $E_k = \sqrt{(\e_{k}-\mu)^2 +
\Delta_{SDW}^2}$
with a gap of $2\Delta_{SDW}$ at the Fermi level. To note, there is a formal 
similarity
of the SDW mean field theory with that of the BCS, in energy spectrum,
gap equation, as well as in the collective modes (it is $2\Delta_{SDW}$ in case
of the
SDW state).
\section{The Coexistence of SDW and SC}

Usually, magnetism and superconductivity are expected to be mutually
exclusive phenomena i.e, they are unlikely to occur simultaneously in the
same compound. Superconductivity (including in the high temperature
superconductors) is known to be due to Cooper pair formation of electrons of
opposite {\it spins} and momenta whereas magnetism requires {\it spin}
 polarisation.
Therefore, naturally one order would inhibit the other. Furthermore,
like superconductivity (electron-electron pairing), the transverse
SDW state is also a result
of condensation of electron-hole pairs of opposite spins but with a momentum
difference of $\vec Q$ between the conjugates. Hence, when any of the orders
(either the SC or the SDW) set in, the FS is instable with respect to that
`condensate state'. In other words, if one of the phases (say SDW)
sets in first,
and exists all over the FS, then there will be no carrier available
to form cooper pairs and hence no superconductivity.
This would also be equally true in case
of the SDW state, had the superconductivity appeared first.
However, in reality we do see the coexistence of the two phases as is already
discussed earlier. In what follows we discuss two different scenarios
where both the
presence of SDW state and SC coexist.

\subsection{The case of an isotropic SDW-state : SC can arise over the SDW backg
round}

If the nesting of the FS is perfect
the entire FS could be isotropically gapped
due to the formation of the SDW state
transforming the system from a metal to an insulator.
In order to build superconductivity over such
an insulating phase one needs to dope the system with charge carriers.
In the event of doping
the system with holes,
or equivalently removal of electrons
from the lower
filled (valence) SDW band,
there will be deviation from perfect nesting of the FS
resulting in a local suppression
of the SDW gap. This is so because, the gain in electronic
energy resulting from the formation of the SDW state is lowered
due to removal of electrons from the valence band.
This local suppression of the SDW gap acts as a
potential well for the injected hole, in which it gets self-trapped
 forming the
so called spin bag. On creating two holes it is energetically
favourable for them to dig a deeper well and stay together provided the
two holes have opposite spins to avoid Pauli exclusion principle.
This is however nothing but
a local Cooper pair and if such bags with two holes of
opposite spins move coherently,
the system will be superconducting. This is the essence of Schrieffer's spin
bag model for high temperature superconductivity \cite{12} and hence is an
example where superconductivitycan arise over the SDW state.

\subsection {Anisotropic SDW state : SC can coexist with SDW}

The above case is however, physically resonable only when the hole
concentration
is very small. But in reality superconductivity in most of the systems
(specially
high $T_c$ systems) appears only after large doping. So, it is likely
that the
SDW state will be completely suppressed in particular directions of the FS
whereas it would still exist in the rest of the FS which still
nests (anisotropic SDW). Such situation may also appear due to particular
topology
of the FS of a system without doping i.e, the system may have nested pieces
of FS only
in certain direction and no nesting in other direction resulting in
an anisotropic SDW state.
In the regions of the FS where the SDW gap vanishes (i.e, the lower and
upper SDW bands merge together), the pairing interaction between
the SDW quasi particles can take
place leading to superconductivity. From a microscopic point of view,
the effective attractive interaction between the SDW quasi particle in
this picture can be rationalised (Ghosh and Sardar \cite{13}) as arising due to
the exchange of the collective modes of the SDW state. In both the
cases discussed above, the origin of superconductivity is fundamentally
different from that in conventional superconductors.
\par In contrast, in the cases where nesting is not perfect either
due to peculiar
topology of the FS or due to doping,
the SDW gap will appear only in the nested part of the
FS allowing for the superconducting instability in the rest
of the FS. But in such a case, the origin of
superconducting pairing may arise be due to any other mechanism
including the BCS phonon exchange \cite{15}. Besides, at least in the case
of high-$T_c$ SCs there are other proposals for SC like
the charge transfer mechanism \cite{16} and the
pair tunneling mechanism \cite{17} etc. However, the nature of SC
pairing and the symmetry of the order parameter in the cuprates
 is still unknown. It was proved by Dzyaloshinskii and Yakovenko
\cite{dy} from a generalised interaction through renormalisation group
equations that the two dimensional Hubbard model could lead to
various instabilities including single superconducting $d_{x^2-y^2}$
pairing, along with possible coherent mixture of charge density wave (CDW)
 and SDW phases. Ruvalds and co-works \cite{ru} also found that $d_{x^2-y^2}$ pa
iring
was favoured for nested fermi surfaces consisting of parallel orbit segments.

The Hamiltonian for the coexistence phase of superconductivity and
the SDW state can therefore be obtained
by adding a pairing interaction
term to equation (6) which in the mean field BCS approximation
can be written as
\begin{eqnarray}
H &=& \sum_{k,\sigma} (\e_k -\mu) C_{k\s}^\dagger C_{k\s} + \Delta_{SDW} \sum_k
( C_{k+Q\uparrow}^\dagger C_{k\downarrow} + h.c.) \nonumber \\
&&
+ \sum_k \D_{sc}(k)  (
C_{k\uparrow}^\dagger C_{-k\downarrow}^\dagger + h.c.)
\end{eqnarray}
where $\D_{sc}(k) =- \displaystyle\sum_{k,k^\prime} V_{k,k^\prime}
<C_{k^\prime\uparrow}^\dagger C_{-k^\prime\downarrow}>$,
$V_{k,k^\prime}$
being the strength of the attractive pairing interaction mediated by some
boson exchange. Depending on the nature of pairing interaction the
superconducting gap function exhibits different types of symmetry structure.
\par In a weak coupling pairing theory, the SC gap parameter
satisfies the equation,
\begin{equation}
\Delta_{sc} (k) = \sum_{k^\prime} V(k,k^\prime)
{{\Delta_{sc}(k^\prime)} \over {2E_{k^\prime}}} \tanh {\beta
E_{k^\prime} \over 2}
\end{equation}
where $E_k = \sqrt{\epsilon_{k}^2 + \Delta_{sc}^2(k)}$ is
the energy of the superconducting quasiparticles.
While $\Delta_{sc} (k)$ is not gauge invariant and
therefore cannot be directly observed, $\mid \Delta_{sc}(k)
\mid^2$ is observable. For a nodeless $\Delta_k (\equiv \Delta_{sc}(k))$,
the state is
non-degenarate and $\mid \Delta_k \mid ^2 $ has the symmetry of
the crystal. However, if $\Delta_k$ has nodes in certain
directions of k, there may be several degenerate pairing
states and $\mid \Delta_k \mid ^2$ need not exhibit the point
symmetry of the crystal --- which is usually referred to as unconvensional
pairing state. More precisely, the symmetry of the order parameter is
determined by the form of the potential $V$ and the band
structure. In the simpliest assumption the band
energies are independent of the direction in the $a-b$ plane and
hence the interaction potential can be approximated as a separable one, i.e,
\begin{equation}
V_{k,k^\prime} = -V \eta_k \eta_{k^\prime}
\end{equation}
where (a) $\eta_k$ = constant, corresponds to an isotropic
s-wave (BCS) symmetry ; (b) $\eta_k$ = f(k), refers to an
anisotropic s-wave symmetry, provided f(k) is a smooth function in the
first Brillouin zone (BZ) and is +ve definite (i.e, nodeless) (c) $\eta_k =
\cos(k_x a) + \cos(k_ya)$, correponds to an extended s-wave ($s^\star$)
pairing symmetry whereas (d) $\eta_k = \cos(k_xa) - \cos(k_ya)$
corresponds to $d_{x^2 - y^2}$ pairing symmetry which also
changes sign (like (c)) but {\it does  not} transform as the
identity under the full crystal group.

Now, in order to study the interplay between the SDW and the
superconducting state (with different possible pairing symmetry
as discussed above) we need to calculate the self-consistent
gap equations for the respective gaps (e.g, SDW $\&$ SC). For this
purpose it is convenient to write the Hamiltonian (7) in
diagonalised form using four component Nambu operator \cite{19} as follows,
\begin{equation}
H_{diag} = \sum_{k} \psi_{k}^\dagger (\epsilon_k \rho_3 \s_3 +
\D_{sc} \rho_3 \s_1 \D_{SDW} \rho_1 \s_3)
\psi_{k}
\end{equation}
where the four component Nambu operators are defined as,
\begin{equation}
\psi_{k}^\dagger = \pmatrix{ c_{k \uparrow}^\dagger & c_{-k \downarrow}
 & c_{k+Q \downarrow}^\dagger  & c_{-k-Q \uparrow}}
\end{equation}
and $\s$, $\rho$ are $4\times4$ matrices defined as,
\begin{equation}
\s_i =\pmatrix{\tau_i & 0 \cr
0 & \tau_i},
\end{equation}
$\tau_i$ being the $2\times 2$ Pauli matrices and $\rho_i$ are given by
\begin{eqnarray}
&&\rho_1 = \pmatrix{ 0 & 0&1&0 \cr 0 & 0 & 0 & 1 \cr 1 & 0 & 0&0 \cr
0 & 1& 0 &0},
\rho_2 = \pmatrix{ 0 & 0&-i&0 \cr 0 & 0 & 0 & -i \cr i & 0 & 0&0 \cr
0 & i& 0 &0} \nonumber \\
&&
\rho_3 = \pmatrix{ 1 & 0&0&0 \cr 0 & 1 & 0 & 0 \cr 0 & 0 & -1&0 \cr
0 & 0& 0 &-1}
\end{eqnarray}

% Such a mean field theory for the coexistence
%of the SDW and the SC phase is well described in
%literature \cite{19}.
The gap equations corresponding
to the order parameters $\D_{SDW}$ and $\D_{sc}$ defined in equations
(6,7) can be calculated  by using equation (10)
which are given below along with the
equation for the total number of charge carriers.
\begin{equation}
\Delta_{SDW} = (-\frac{U}{4}) \sum_{k, i=1,2} (-1)^i
\frac{\Delta_i (k)}{E_{i}(k)} \tanh \frac{\beta E_i(k)}{2}
\end{equation}
and
\begin{equation}
\Delta_{sc} (k) = (\frac{1}{4}) \sum_{k^\prime i=1,2} V_{k,k^\prime}
 \frac{\Delta_i (k^\prime)}{E_i(k^\prime)}
\tanh \frac{\beta E_i(k^\prime)}{2}
\end{equation}
\begin{equation}
n = 1 - \frac{1}{2} \sum_{k,i=1,2} \frac{(\epsilon_k -\mu)}{E_{i}(k)}
\tanh \frac{\beta E_{i}(k)}{2}
\end{equation}
with
\begin{eqnarray}
E_{i}(k)
&=&\sqrt{(\epsilon_{k} -\mu)^2 +\Delta_{i}^2(k)}
\rm ~and~
\nonumber  \\
& &
\Delta_i(k) = (\Delta_{sc}(k) - (-1)^i \Delta_{SDW}).
\end{eqnarray}

The equations (14-16)
are coupled integral equations in the sense that $\Delta_{sc}
\equiv \Delta_{sc} (\Delta_{sc}, \Delta_{SDW}, n)$,
$\Delta_{SDW} \equiv \Delta_{SDW}(\Delta_{sc}, \Delta_{SDW}, n)$ and
so is $n \equiv n(\D_{sc}, \D_{SDW}, n)$.
The self-consistent solutions of these three equations are obtained
 numerically.
Before the  numerical results are presented it is worth pointing
out that it appears from equations (14-17) as if there exists two
effective order parameters $\Delta_{1,2} = \Delta_{sc} \pm \Delta_{SDW}$
indicating
that these can interfere with each
other either destructively or constructively. In striking
contrast, in the coexistent phase
of the charge density wave (CDW) and superconductivity \cite{20}
no show such interference is obtained where the quasi particle energy
spectrum has an effective gap $
\sqrt{\D_{CDW}^2 + \Delta^2}$ , $\D_{CDW}$ being the CDW order parameter.
Such a difference in behaviour between the SDW and CDW coexistent with SC
may be attributed to the effect of interference
between the order parameters of the SDW and SC states both of which
involve the pairing of up and down spin quasi-particles, i.e, the
electron-hole and electron-electron
pairings respectively. Therefore, the electrons with the same
spins are likely to compete for both the processes, thereby
giving rise to interference.
\section{Results and Discussion}
The interplay between the SC and SDW state is mainly studied by
selfconsistently solving the gap equations (14 - 16) with fixed
set of parameters $U = 1, ~ V = 1.5,$ the cut-off energy for SC
$\O = 0.8$, all are in units of the hopping intgral $t$. Two distinct
set of results are obtained, (i) for a fixed temperature (5 K) the
amplitudes of the gap functions (SDW $\&$ SC) are obtained for
different pairing symmetry as a function
of band filling ($n$) and (ii) temperature variation of the
respective order parameters (for different pairing symmetry)
are obtained
for different (fixed) band fillings. In the former case, the thermal
variation of the chemaical potential, $\mu(T)$ is not important
whereas for the later the thermal variation of $\mu(T)$ is taken care
such that given value of $n$ differs from that of the calculated ones
at best at the fifth decimal point.
Furthermore, in computing the self-consistent solutions of equations
(14 - 16) for different pairing symmetries of the superconducting
state, specific forms of the pairing potential and the corresponding
order parameters are also assumed. Following
the form of the potential in equation (15) we assume the form of the
order parameters, for example, for
the extended $s$ ($s^\star$) and $d$ wave SC as,
$\D^{s(d)} = \D^0 (\cos k_x \pm \cos k_y)$.

Now we  will discuss the results of our calculation. In fig. 1. we have plotted
the isotopic s-wave SDW gap parameter ($\Delta_{SDW}$)
and isotopic s-wave SC gap parameter ($\Delta_{SC}$) evaluated at T = 5K
vs. hole
concentration $x=1-n$. Clearly the
figure demonstrates the strong competition between the two orderings namely SDW
and
SC emphasizing the coexistence of SDW state and Superconductivity at low hole
concentration $x <0.23 $ .
As we increase the hole concentration first SC gap and then the SDW gap
goes to zero. As a result $0.23<x<0.25$ we observe a pure SDW phase. When
the hole concentration is further increased we obtained pure superconducting
phase for $0.25<x<0.5$. This reentrance of superconducting phase
with respect to hole concentration is quite a new phenomena and to the best
of our knowledge no such theoretical demonstartion exists.
%the SDW gap suddenly
%drops to zero and we obtain a pure superconducting phase with $0.2<x<0.5$.

In fig.2. we have shown the variation of SDW gap function
and amplitude of extended s-wave ($s^\star$) SC gap function $\Delta_{SC}^0$ 
evaluated
at 5K with the hole
concentration $x$ for the same set of parameters as in Fig. 1.
Here there is  no such coexistence of SDW
and SC phase. In fact SDW phase is dominant in the low hole concentration
region where as SC phase is observed at large hole concentration.
More precisely
SDW gap shows a maxima at half filling ($x=0$), decreases with the increase of x
and ultimately vanishes at around x=0.5 where as extended s-wave SC gap
is seen at around
$x > 0.65$ with a maxima around $x=0.8$.  Comparing the figure 1 and 2 we find 
that
superconducting ordering suppresses the SDW ordering and vice versa 
although for isotopic 
s-wave
pairing one finds the coexistence of SDW and SC phase.

In fig. 3. we have presented the scenario of d-wave superconductivity with the
SDW. In the
undoped (x=0) state the system is an insulating antiferromagnet with a
SDW gap.
As the system is doped with holes the antiferromagnetic ordering is suppressed
and the  system becomes
superconducting in the region $0.4\geq x \geq 0.25$. When $x>0.4$ superconductiv
ity disappears
and we obtain a reentrant
SDW phase. However, in this region of hole concentration ($x\ge 0.4$) 
the SDW gap parameter is small. Possibly
the quantum fluctuation which has been neglected in this mean  field calculation
 will destroy
the SDW ordering.

\par The SDW gap being isotropic $s$ wave it has always maximum value at the
half-filling because the density of states (DOS) is maximum at the
Fermi lavel (half-filing) only. Since the Fermi surface is determined by the
zero energy contour of $\epsilon_k = -2t(\cos k_xa + \cos k_ya)= 0$, the factor
$(\cos k_x a + \cos k_y a)$ is therefore smaller, for a square lattice, close
to half-filling (as $n =1$ corresponds to the position of the Fermi Surface)
and has larger value away from half-filling. Therefore, the self consistent
solutions of the amplitudes of the extended $s$ ($d$) wave superconducting
gap will always be minimum (maximum) at half-filling whereas maximum (minimum)
at zero filling. This would qualitatively explain the band filling dependence
of the extended $s$ wave and $d$ wave superconducting gaps (as is presented
in Figures 1 -- 3). The presence of the SDW order makes the situation
complicated (and probably more realistic for the systems like the
 cuprates). The magnitudes
of the SDW gap is maximum at half-filling not only because of the large
van Hove DOS but also due to the fact that the nesting of the FS is strong
close to half-filling. This results in the appearence of the SDW phase at
lower doping concentrations ($x)$ irrespective of the nature of 
the superconducting
pairing (cf figures 1-4). Considering the case of interplay between the
SDW and $s$ wave superconductivity we therefore find that both the SDW $\&$ SC
states are degenerate at $x=0$. However, the SDW phase is more stable than the
SC phase although the interaction strength is smaller ($U<V$) due to strong
nesting effect. Contribution to superconducting phase at half-filling is
probably from the non-nested parts of the FS. However, with doping the nesting
of the FS gets affected and superconducting phase becomes stable suppressing
the SDW phase (cf Fig. 1, Fig. 3). 
In case of extended $s$ wave pairing therefore, no
strong interplay is found as expected (cf. Fig.2). In case of $d$- wave
pairing (cf. Fig. 3)
however, the amplitude of the SDW order parameter is large at $x=0$ compared
to that for $s$ wave pairing scenario. 
A closer comparision of the figures 1 $\& $ 3
will also reveal that the maximum value of $d$ wave SC gap is approximately
three times larger than the maximum value of isotropic $s$ wave SC gap (within
the same set of all other parameters). This is because for isotropic
$s$ SC wave pairing, both the pairing strengths (SC as well as the SDW)
 are onsite in nature and
will be most effective only at the half-filling due to symmetry reason (as
 discussed earlier). As a result both the orders will inhibit each other.
In contrast to isotropic $s$ wave scenario, for $d$ wave pairing the SC phase
is however pushed towards higher doped region
 (unlike being degenerate at $x = 0$).
This is due to the fact that the more extended
 d-wave pair wave function
is more efficient in avoiding the
pair breaking effect of the staggered local field.   
In the conserving fluctuation exchange approximation (FLEX) \cite{18} which are
in good agreement with available monte carlo data at higher temperatures
showed that when the two dimensional Hubbard model was doped beyond a
critical doping the leading instability changes from the SDW channel to
$d_{x^2-y^2}$ pairing channel. Therefore, our results in the figure 3
from a mean field calculation agrees qualitatively with that of FLEX
\cite{18} approximation calculation.
Furthermore we observe that the gap
function for d-wave superconductors decreases
sufficiently rapidly from its maximum value with hole
concentration in comparison
with isotopic s-wave or extended s-wave superconductivity.
Finally, in contrast to  
s-wave pairing d-wave superconductivity does not show any
coexistence of SDW and SC phase.

In figure 4. we have plotted the SDW gap and an admixture of $s$ and
$d$ wave pairing (s+d) superconductivity (with SC gap having $80 \%$
 d-wave
contribution and $20 \%$ extended s-wave contribution) vs $x$.
Here we find that the superconducting phase appears only within
very narrow range of hole (almost does not appear) concentration.
Also the highest value of the SC gap amplitude is much smaller
than that seen for the pure $s$ or $d$ wave picture. This is probably
an indication of the fact that the superconducting state with mixed $s
+d$ symmetry of the order parameter
is stable only in presence of orthrhombic distortion not in tetragonal phase
\cite{21}. We shall present results in details of the interplay of the SDW
state with complex ($s + id$) SC pairing in our next publications.

\par Fig. 5 $\&$ 6 demonstrates the thermal variation of the $s$-wave
SC gap and the SDW
gap for different hole concentrations. In Fig. 5, (for given set of parameters
as earlier) for the half-filled case ($x=0.0$) the SC gap opens up at higher
temperatures ($T_c$ =161K), grows with the lowering in temperature but drops
to zero at 100K till 75 K. On the other hand, the SDW gap opens up at 100K
and grows with the lowering in temperature but its growth being arrested at
75 K (due to the re-entrance of the SC state at the same temperature) on
further lowering the temperature. Away from half-filling, at $x = 0.2$, the SC
$T_c$ is reduced drastically (to $\sim 85$K) and the SC-gap drops to zero
below $T_c$ at 28 K and reappears at 21 K. The SDW gap (at $x = 0.2$) grows
like a first order transition at around 28 K (where SC-gap drops to zero) and
grows with lowering in temperature which is again being arrested with the
reentrance of SC below 21 K.
% At $x =0.3$ no  SDW state has been observed but
%the SC gap is almost independent of temperature at lower temperature and
%suddenly drops to zero at 21 K.
Therefore, Fig. 5 demonstrates very strong interplay and co-existence 
between the SDW and the $s$-wave state. To be noted the
strong interplay between the two ordered phases that includes their coexistence
as well as re-entrance, is maximum at half-filling.

\par While in Fig. 5 there exists coexistence as well as re-entrance of
different
parameters, Fig.6 demonstrates another particular aspects of their interplay.
The hole concentration here is $x = 0.1$ and one finds that for certain range
of thermal region, around 48 to 65 K none of the order parameters are stable.
In other words, within this thermal regime both the order parameters
precipetously either goes to zero or a finite value alternatively
for a very small change in temperature. In the rest of the
thermal regime it retains the essential features of the earlier figure (Fig.5)
as mentioned earlier. The arrest of the growth of the SDW gap at lower
temperature with the re-entrance of the SC gap can be understood as follows.
The SDW gap (in the mean field treatment) being a very sensitive function
of the density of state available at the FS, the opening up of the SC gap
takes away large density of states and hence the SDW gap is suppressed at
lower temperatures. On the other hand, the rentrance as well as the thermally
fluctuating natures are due to typical interefernce between the two order
parameters which are described to some extent analytically by the equation
(17). Therefore, the present paper presents as extensive study on the
influence of the SDW order on the SC pairing symmetry. We would like to further
point out that the figures 1 $\&$ 5 have resemblence with the spin
susceptibility data of borocarbides \cite{22}, indicating an interplay of
magnetism and superconductivity in these systems. \\

\noindent{\bf Acknowledgement} One of us (HNG) would like to acknowledge the
hospitality of Prof. Dr. Peter Entel, through a grant from SFB166, which
made it possible for him to visit the Gerhard-Mercator University at
Duisburg, Germany where a part of the present work had been carried out. We
would also like to thank him for many useful discussions on the subject. One
of us (HNG) thanks the Brazilian agency FAPERJ for providing the financial
support (Project No. E-26/150.925/96-BOLSA).

\end{multicols}
\begin{figure}
\epsfxsize=4.5truein
\epsfysize=4.5truein
\begin{center}
\leavevmode
\caption{Amplitudes of the SDW and the $s$-wave SC gap
evaluated at 5K, as a function of the hole concentration ($x$).}
\end{center}
\end{figure}

\begin{figure}
\center
\epsfxsize=4.5truein
\epsfysize=4.5truein
\leavevmode
\caption{The SDW and extended $s$-wave gap amplitudes
evaluated at 5 K is plotted as a function of the hole concentration ($x$).}
\end{figure}

\begin{figure}
\center
\epsfxsize=4.5truein
\epsfysize=4.5truein
\leavevmode
\caption{Gap amplitudes of the SDW and the $d_{x^2-y^2}$ SC gap a
s a function
of hole concentration ($x$).}
\end{figure}

\begin{figure}
\center
\epsfxsize=4.5truein
\epsfysize=4.5truein
\leavevmode
\caption{The SDW and SC gap in the mixed $(s+d)$ symmetry
is plotted as a function of the hole concentration.}
\end{figure}

\begin{figure}
\center
\epsfxsize=4.5truein
\epsfysize=4.5truein
\leavevmode
\caption{Thermal variation of the isotropic $s$-wave SC gap
and the SDW gap (in units of $t$) for various fixed hole concentrations
$(x)$.}
\end{figure}

\begin{figure}
\center
\epsfxsize=2.5truein
\epsfysize=2.5truein
\leavevmode
%{\epsffile{fig5.eps}}
%\input{shn6}
\caption{Temperature variation of the isotropic $s$-wave
SC gap and  the SDW gap for $x = 0.1$. The thermally instable region is
worth noting.}
\end{figure}

\begin{references}
\bibitem{1} T. Ishiguro, K. Yamaji, ``Organic superconductors" ;
Springer series in Solid state Sciences, Vol. {\bf 48}, p-99, Berlin,
Heidelberg, New York ; Springer 1990.
\bibitem{2} For a recent review in organic superconductors see,
H. Mori, Int. Jr. Mod. Phys. B {\bf 8}, 1 (1994) and L. N. Bulaevski,
Adv. Phys. {\bf 37} (1988) 443.
\bibitem{3} M. Kato and K. Machida, Phys. Rev. B {\bf 37} (1988) 1510 ;
C. Geibel et al, Z. Phys. B {\bf 86}, 161 (1992) ; R. Caspary et al.,
Phys. Rev. Lett. {\bf 13} 2146 (1993).
\bibitem{4} D. W. Murphy et al., Phys. Rev. Lett. {\bf 58}, 1888 (1987) ;
J. W. Mills et al., Jr. Magn. $\&$ Magn. Mat. {\bf 67} L 139 (1987).
\bibitem{5} R. Nagarajan et al., Phys. Rev. Lett. {\bf 72}, 274 (1994) ;
R. J. Cava et al., Nature {\bf 367}, 146 (1994) ; H. Eisaki et al.,
Phys. Rev. B {\bf 50}, 647 (1994).
\bibitem{6} K. Kitazawa et al., Jpn. Jr. Appl. Phys. {\bf 26}, 339 (1987).
\bibitem{7} Z. -X. Shen et al., Science {\bf 267}, 343 (1995) ; H. Ding
et al., Phys. Rev. Lett. {\bf 76} (9) 1533 (1996); S. LaRosa, I. Vobornik, 
F. Zwick, H. Berger, M. Grioni, G. Margaritondo, R. J. Kelley, M. Onellion
ans A. Chubukov, Phys. Rev. B {\bf 56} R525 (1997).
\bibitem{8}  G. Shirane et. al. Phys. Rev. Lett {\bf 59}, 1613 (1987);
J. Transquada et. al. Phys.Rev. Lett. {\bf 60}, 159 (1988).
\bibitem{9} K. B. Lyons et. al. Phys. Rev. B {\bf 37}, 2353 (1988);
Phys. Rev. Lett. {\bf 60}, 732 (1988).
\bibitem{10} P. W. Anderson, Science {\bf 235}, 1196 (1985).
\bibitem{11} P. Monthoux and D. Pines, Phys. Rev. B {\bf 47}, 6069 (1993)
and references therein.
\bibitem{12} J. R. Schrieffer, X. G. Wen, Phys. Rev. Lett. {\bf 60},
944 (1988) ; Phys. Rev. B {\bf 39}, 11663 (1989).
\bibitem{13} Haranath Ghosh and M. Sardar, Physica C {\bf 246}, 335 (1995).
\bibitem{14} J. Ma et al., Science {\bf 267},865 (1995) ;
M. Ichioka, E. Enomoto, N. Hayashi
and K. Machida, Phys. Rev. B {\bf 53} (5) 2233 (1996); Haranath Ghosh,
Europhys. Lett. {\bf 43}, 707 (1998); Haranath Ghosh,
Phys. Rev. B {\bf 59}, 3357 (1999).
\bibitem{15} W. E. Pickett, Rev. Mod. Phys. {\bf 61}, 433 (1989).
\bibitem{16} C. M. Verma, S. Schmitt-Rink and E. Abrahams, Solid State Comm.
{\bf 62}, 681 (1987).
\bibitem{17} Sudip Chakraborty et al., Science {\bf 261}, 337 (1993).
\bibitem{dy} I. E. Dzyaloshinskii and V. M. Yakovenko, Zh. Eksp. Teor. Fiz.,
{\bf 94}, 344 (1988).
\bibitem{ru} J. Ruvalds, C. T. Rieck, S. Tewar, J. Thoma and A. Virosztek
, Phys. Rev. B {\bf 51}, 3797 (1995).
\bibitem{18} D. J. Scalapino, Phys. Rep. {\bf 250} 329 (1995).
\bibitem{19} K. Machida, J. Phys. Soc. Jpn. {\bf 50} 2195 (1981) ;
S. N. Behera and S. Bhattacharya, Physica C {\bf 167}, 112 (1990).
\bibitem{20} S. N. Behera and S. G. Mishra, in ``Theoretical and experimental
aspects of valance Fluctuations and Heavy Fermions", Eds. L. C. Gupta and
S. K. Mallik (Plenium Press, N Y) p-265 (1987) ; S. N. Behera and
Haranath Ghosh, Z. Phys. B {\bf 95}, 275 (1994).
%\bibitem{21} M. Inui and S. Donic, Phys. Rev. B {\bf 37} 2320 (1988).
\bibitem{21} C. O'Donvan and J. P. Carbotte, Physica C {\bf 252} 87 (1995) ;
Y. Ren et al., Phys. Rev. B {\bf 53}, 2249 (1996) ; K. A. Musaelian et al.,
Phys. Rev. B {\bf 53}, 3598 (1996).
\bibitem{22} L. C. Gupta, Physica B {\bf 223 - 224}, 56 (1996).
\end{references}
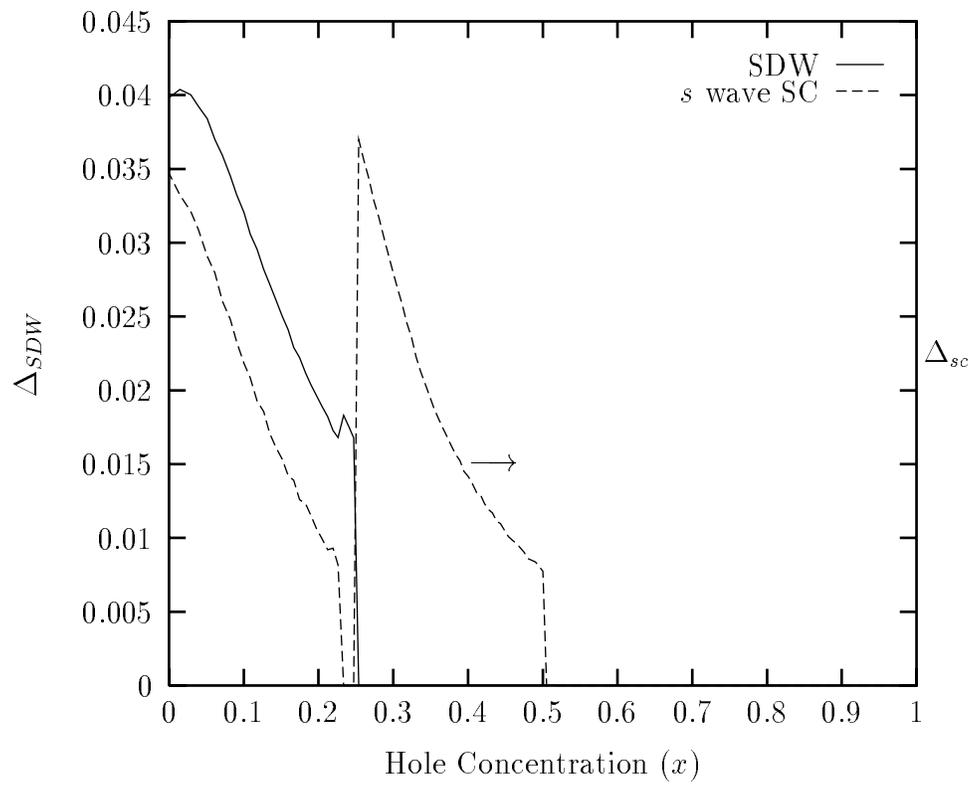
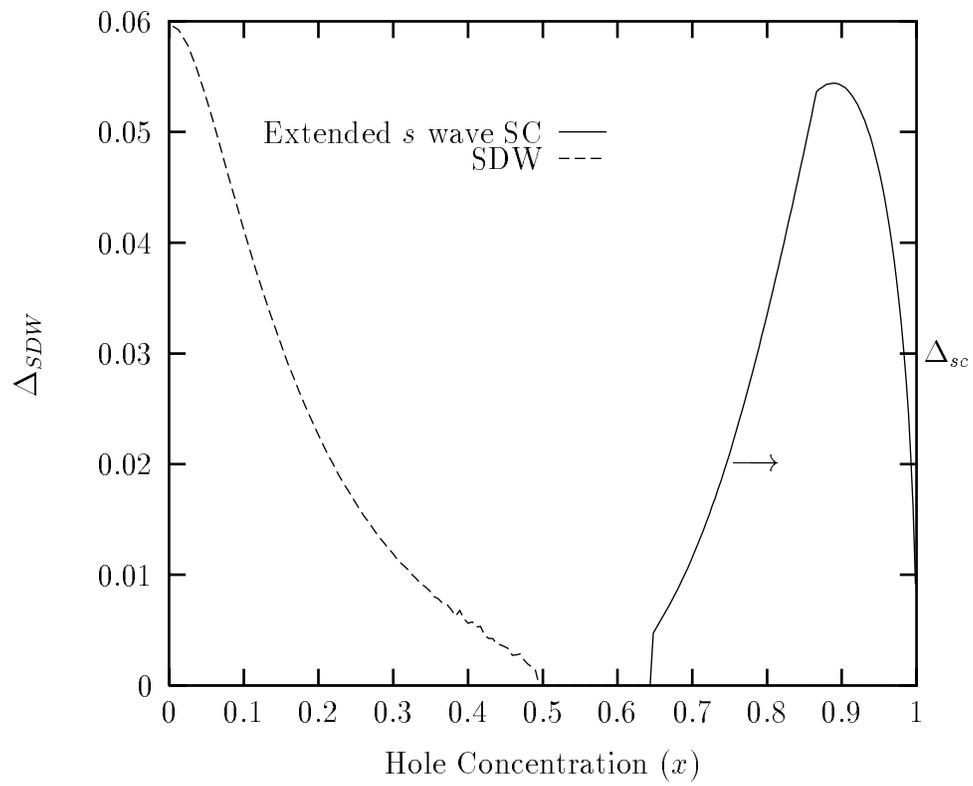
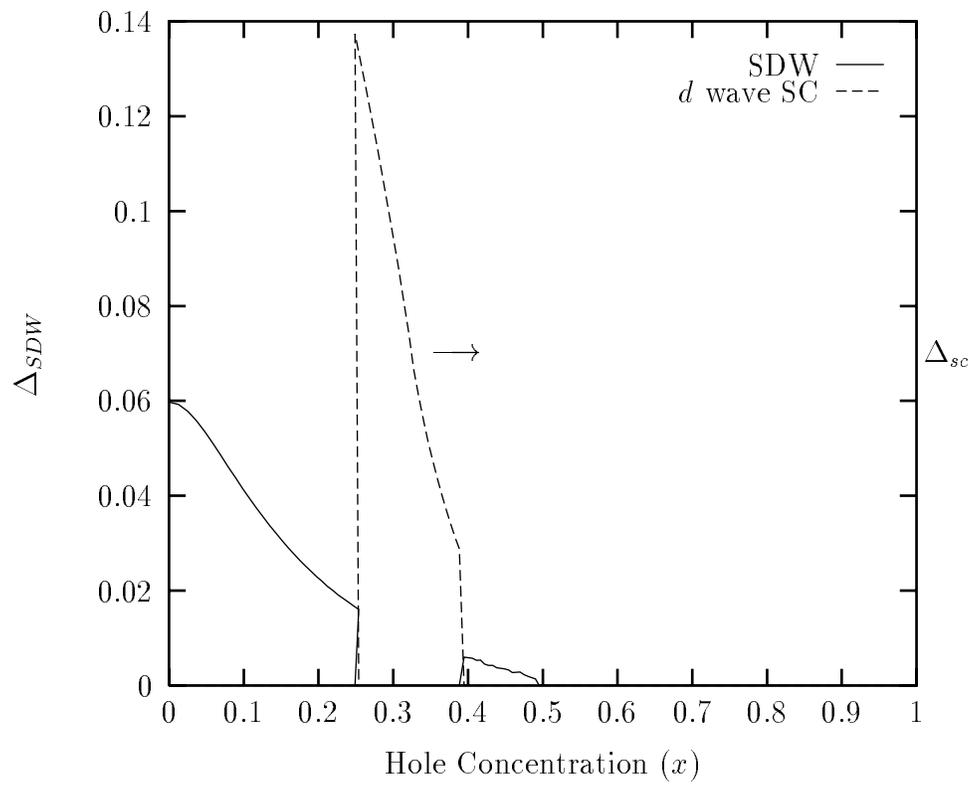
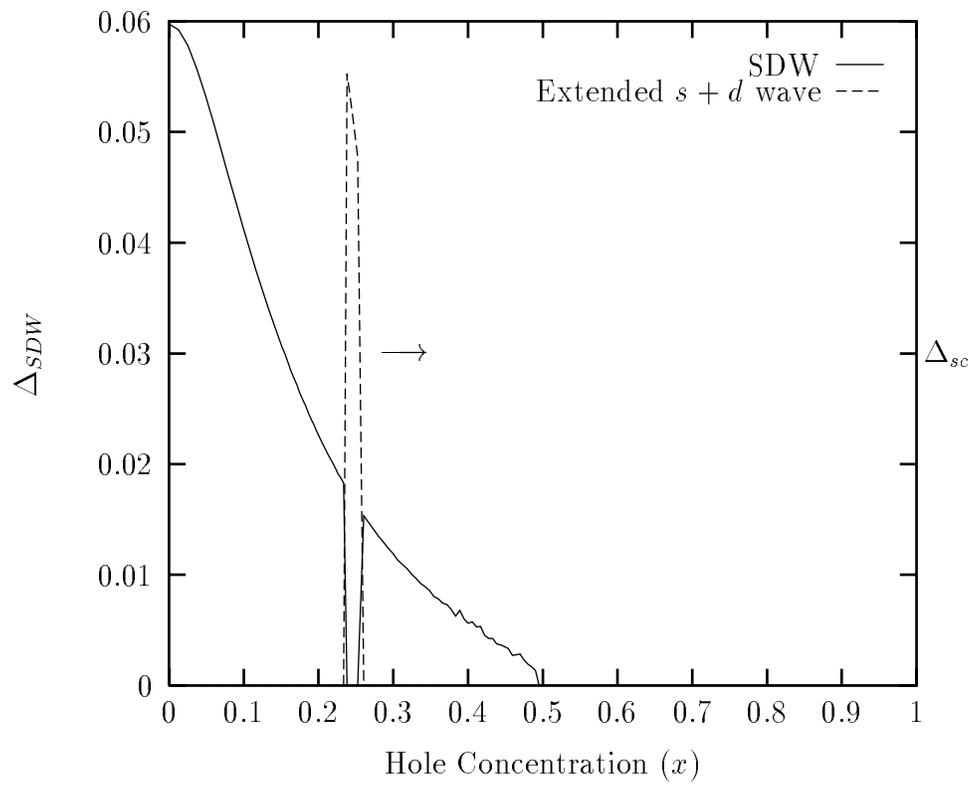
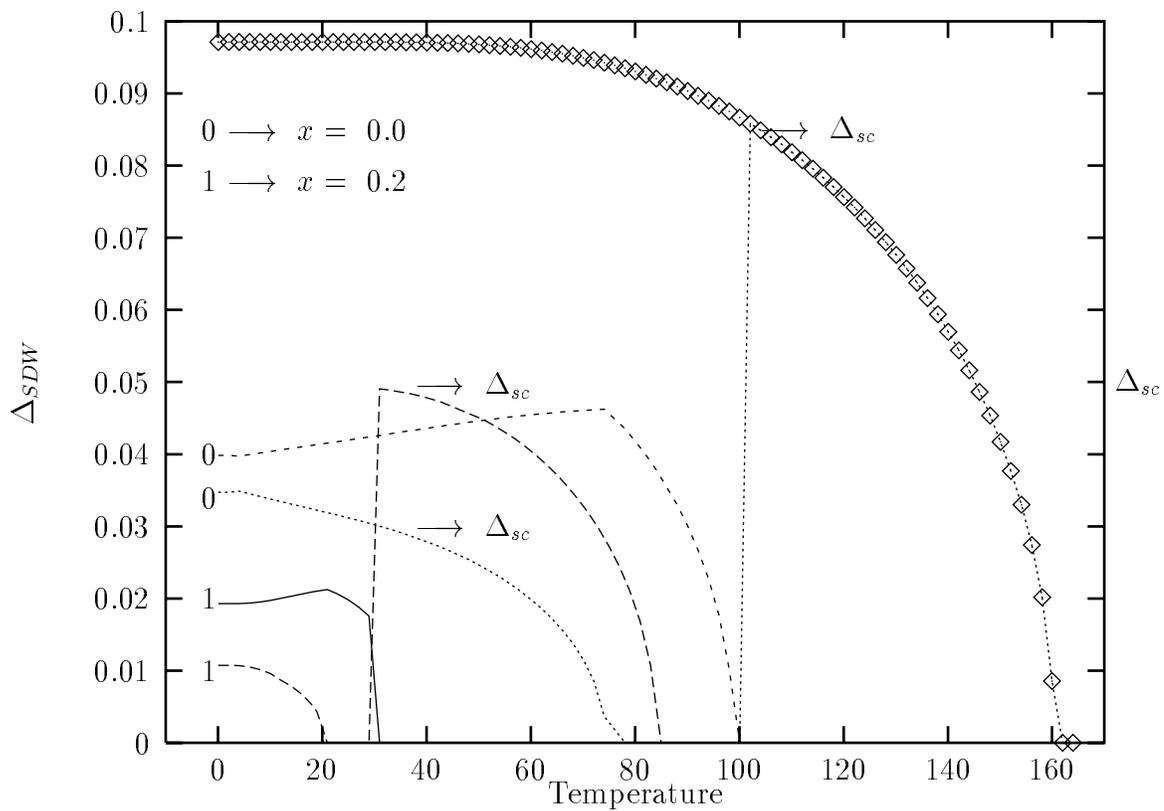
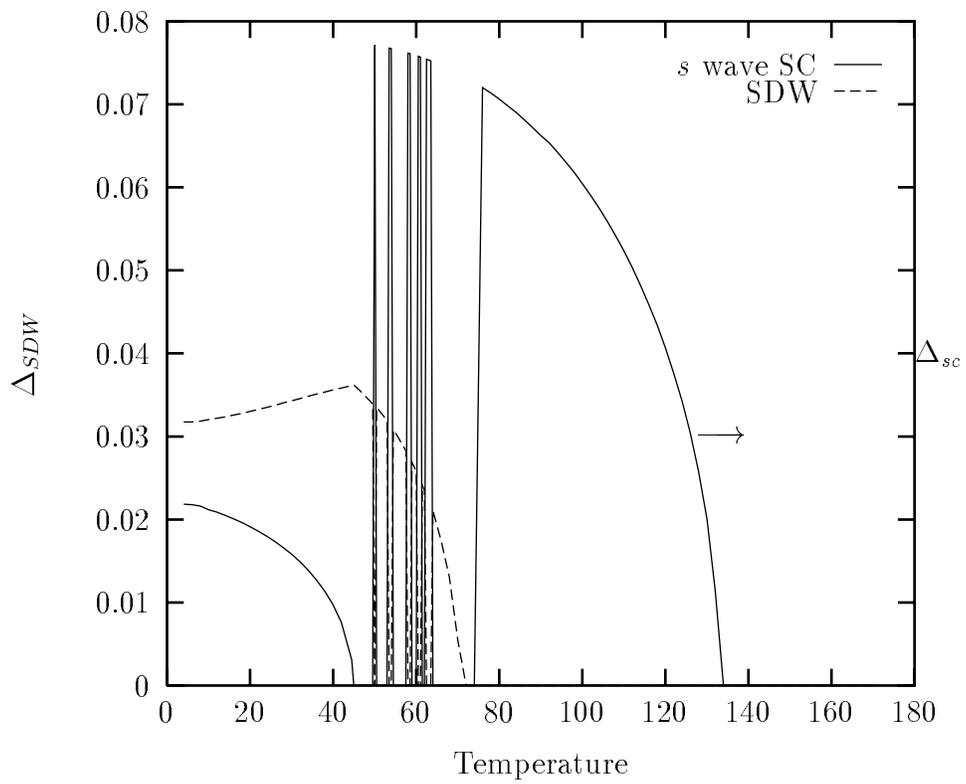
\end{document}